\documentclass[twocolumn,showpacs,preprintnumbers,amsmath,amssymb,superscriptaddress,prl]{revtex4}

\usepackage{graphicx}

\begin{document}

\title{Observation of a Fractional Quantum Hall State at $\nu$=1/4 in a Wide GaAs Quantum Well}

\author{D.R. Luhman}
\affiliation{Department of Electrical Engineering, Princeton University, Princeton, NJ 08544}
\author{W. Pan}
\affiliation{Sandia National Laboratories, Albuquerque, NM, 87195}
\author{D.C. Tsui}
\affiliation{Department of Electrical Engineering, Princeton University, Princeton, NJ 08544}
\author{L.N. Pfeiffer}
\author{K.W. Baldwin}
\author{K.W. West}
\affiliation{Bell Labs, Lucent Technologies, 700 Mountain Avenue, Murray Hill, New Jersey 07974}

\date{\today}

\begin{abstract}

We report the observation of an even-denominator fractional quantum Hall (FQH) state at $\nu=1/4$
in a high quality, wide GaAs quantum well. The sample has a quantum well width of 50 nm and an
electron density of $n_e=2.55\times10^{11}$ cm$^{-2}$. We have performed transport measurements at
$T\sim35$ mK in magnetic fields up to 45 T.  When the sample is perpendicular to the applied
magnetic field, the diagonal resistance displays a kink at $\nu=1/4$. Upon tilting the sample to an
angle of $\theta=20.3^o$ a clear FQH state at emerges at $\nu=1/4$ with a plateau in the Hall
resistance and a strong minimum in the diagonal resistance.
\end{abstract}


\maketitle

Interest in the even-denominator fractional quantum Hall (FQH) state at $\nu=5/2$ in the first
excited Landau level continues to remain high over twenty years after its discovery\cite{willett}.
Generally believed to be due to the $p-$wave pairing of composite
fermions\cite{MooreRead,pfaff,morf,Scarola}, the quasi-particle excitations of this state are
thought to obey nonabelian statistics and thus may be relevant to fault-tolerant, topological
quantum computing schemes\cite{kitaev,TQC}.

To date, observations of even-denominator FQH states have been rare beyond the $\nu=5/2$ state in
single-layer systems\cite{NewPanPRB}. In particular, experimental evidence for a FQH state at
$\nu=1/2$, the lowest Landau level counterpart of the $\nu=5/2$ state, does not exist, although
previous theoretical work has suggested that it may form in thick two-dimensional electrons systems
(2DES)\cite{park}.

In bilayer systems, however, the situation is different. The presence of two nearby interacting
electron layers introduces an additional degree of freedom which can allow the formation of a FQH
state at $\nu=1/2$. Observations of such a state at $\nu=1/2$ have been made in both double quantum
wells\cite{Eisenstein} and  wide single quantum wells (WSQWs)\cite{Suen,Suen94}. In both of these
cases the $\nu=1/2$ state has been shown to have a large overlap with the so-called $\{331\}$ wave
function\cite{SongHe}. Originally proposed by Halperin\cite{HaperinHPA} to describe two-component
FQH states, the $\{331\}$ wave function can also be characterized as a $p-$wave pairing state,
although with albelian statistics\cite{HalperinSS,ReadRezayi,ReadGreen}. A crude way to interpret
the $\{331\}$ wave function, or in general any $\{nnm\}$ wave function, is to consider two electron
layers each with a filling factor of $\nu^*=1/n$. The electrons in each layer are bound to
correlation holes in the other, represented by a filling factor of $1/m$. Together the filling
factor of the entire system is $\nu=2/(n+m)$\cite{perspectives}.

Beyond the $\{331\}$ state the model should generalize to other even-denominator states. For
example, both the $\{771\}$ and $\{553\}$ wave functions would be possible candidates to describe a
FQH state at $\nu=1/4$. In contrast to $\nu=1/2$, relatively little theoretical work has been done
concerning a FQH state $\nu=1/4$ and an experimental observation of this state has yet to be
reported. On the one hand, an observation of the $\nu=1/4$ state would be demanding experimentally,
including a high mobility 2DES and ultra high magnetic fields for high density samples. On the
other hand, an observation of the a FQH state at $\nu=1/4$ would not only be of interest by itself,
but may also further elucidate the still enigmatic $\nu=1/2$ state in WSQWs.

In this Letter, we report the observation a FQH state at $\nu=1/4$ in a high quality GaAs quantum
well of width $L=50$ nm at $T\sim35$ mK  using a 45 T magnet. When the quantum well is oriented
perpendicular to the applied magnetic field, a small kink is seen in the diagonal resistance
($R_{xx}$). When the sample is tilted to an angle of $\theta=20.3^o$, the kink in $R_{xx}$ develops
into a strong minimum and a plateau appears in $R_{xy}$, clearly demonstrating a FQH state at
$\nu=1/4$.

\begin{figure*}[t]
\resizebox{6.8 in.}{!}{\includegraphics{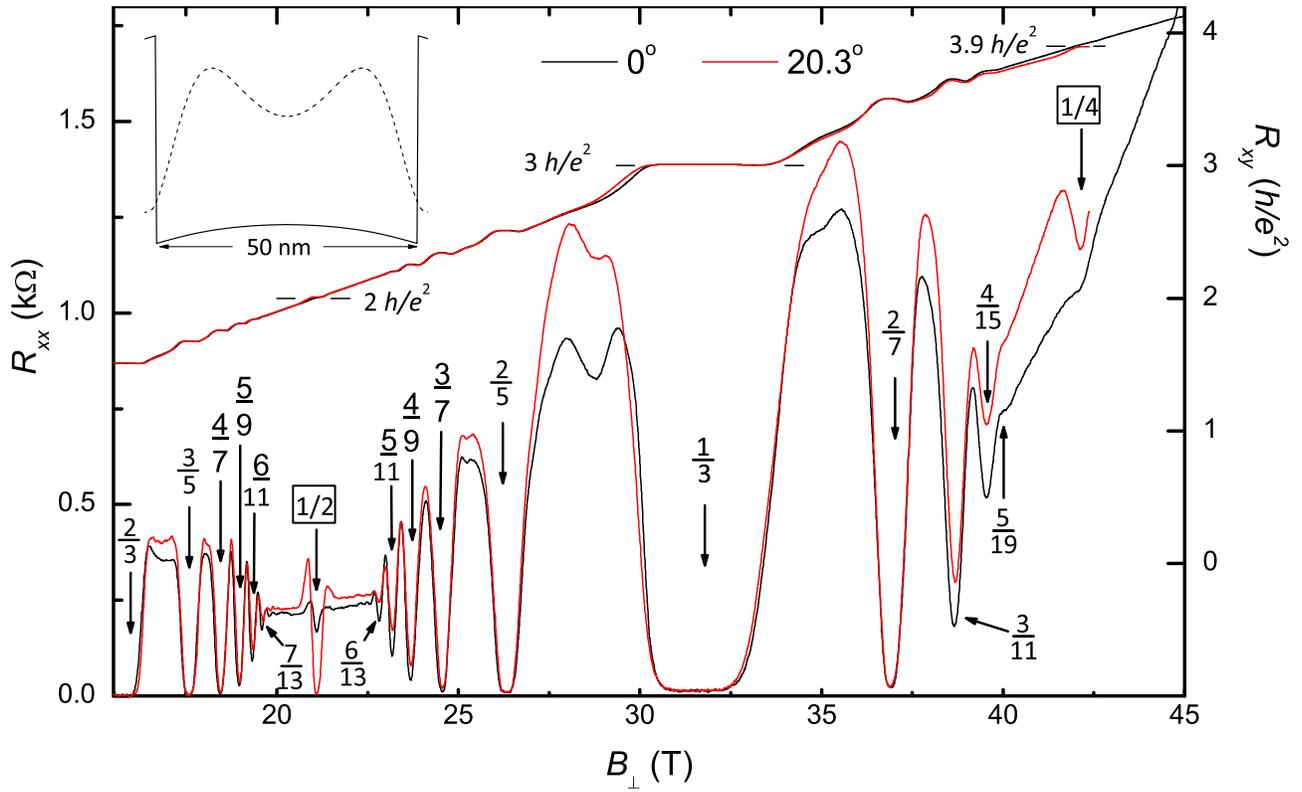}}

\caption{$R_{xx}$ and $R_{xy}$ as a function of the perpendicular component of magnetic field,
$B_{\perp}$, for $\theta=0^o$ and $20.3^o$ at $T\simeq35$ mK. Many of the odd-denominator FQH
states are labeled. The filling factors $\nu=1/2$ and $\nu=1/4$ are also marked. The horizontal
lines at $R_{xy}=2h/e^2$, $3h/e^2$ and $3.9h/e^2$ highlight the plateaus at those values. In the
inset in the upper left corner we show the zero-field, self-consistently calculated results of the
electronic potential (solid line) and the electron distribution across the well (dashed line).
}\label{fig:Fig1}
\end{figure*}

Our sample is a Al$_{0.24}$Ga$_{0.76}$As/GaAs/Al$_{0.24}$Ga$_{0.76}$As modulation doped quantum
well. The width of the well is $L=50$ nm and the two doping layers are symmetrically positioned at
a distance of 80 nm from each side of the quantum well. The specific piece used for the measurement
was a 5 mm $\times$ 5 mm square with eight equally spaced ohmic contacts positioned around the
perimeter. The sample was cooled from room temperature to 4 K over the course of 90 minutes while
being exposed to light from a red light-emitting-diode. The sample was then cooled further to a
base temperature of $T\sim35$ mK in the mixing chamber of a dilution refrigerator. The electron
density of the sample was $n_e=2.55\times10^{11}$ cm$^{-2}$ and the mobility was $\mu\sim10^7$
cm$^2$/Vs. The density of the 2DES varied slightly ($\sim5\%$) depending on the cooling and
illumination conditions. Transport measurements were performed using standard lock-in amplifier
techniques with a low frequency ($\sim7$ Hz) excitation current of 50 nA and negligible heating
effects. All the measurements were carried out using the Hybrid magnet facility at the National
High Magnetic Field Laboratory in Tallahassee, FL, allowing us to reach a maximum magnetic field of
45 T. Tilt measurements were done with an \textit{in situ} rotator and the angle ($\theta$) was
determined by aligning well developed quantum Hall states using $B_{\perp}=B\times\cos\theta$ where
$B$ is the total applied magnetic field and $B_{\perp}$ is the component perpendicular to the plane
of the sample.

\begin{figure}[t]
\resizebox{3.5 in.}{!}{\includegraphics{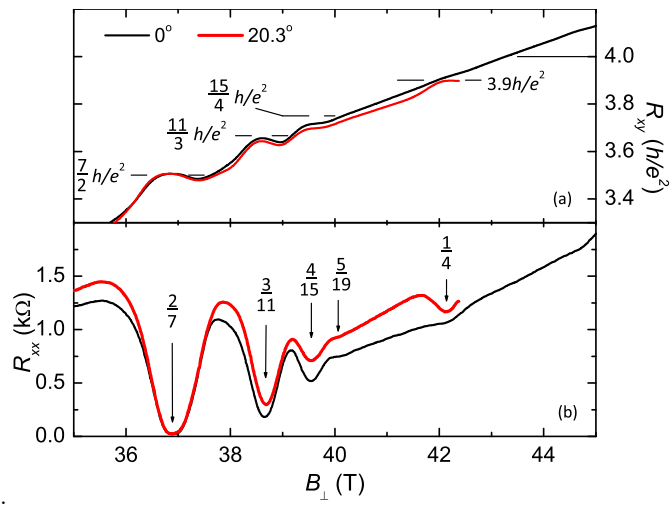}}

\caption{$R_{xx}$ and $R_{xy}$ as a function of $B_{\perp}$ near $\nu=1/4$ for $\theta=0^o$ and
$20.3^o$. The odd-denominator states and the FQH state at $\nu=1/4$ are indicated by arrows. The
expected values of the plateaus in $R_{xy}$ are marked with horizontal lines. The actual value of
the plateau $R_{xy}=3.9h/e^2$ at $\nu=1/4$ is also shown. }\label{fig:Fig2}
\end{figure}

In Fig.~\ref{fig:Fig1} we display $R_{xx}$ and $R_{xy}$ as a function of $B_{\perp}$ for
$\theta=0^o$ and $\theta=20.3^o$. The high quality of the sample is evident from the large number
of FQH states surrounding $\nu=1/2$; clear minima are seen to $\nu=7/13$ and $6/13$ on the low
field and high field side of $\nu=1/2$ respectively. Furthermore, a FQH state is present at
$\nu=1/2$ with both a minimum in $R_{xx}$ and a weak plateau in $R_{xy}$. This FQH state becomes
fully developed at $\theta=20.3^o$ with $R_{xx}$ vanishingly small and $R_{xy}$ precisely quantized
at $2h/e^2$. Moving beyond $\nu=1/3$, FQH states are visible at $\nu=2/7$, 3/11, 4/15, and 5/19,
their strength weakening as $\nu=1/4$ is approached. In the absence of tilt, a kink is noticeable
in $R_{xx}$ at $\nu=1/4$ along with a very subtle deviation from the classical Hall slope in
$R_{xy}$. With the sample tilted to $\theta=20.3^o$ the strength of the odd denominator FQH states
at $\nu=1/3$ and $2/7$ varied little, while the states at $\nu=3/11$, 4/15, and 5/19 weakened
slightly. In contrast, at $\nu=1/4$ the kink in $R_{xx}$ has developed into a well defined minimum
on top of the rising background and a plateau has emerged in $R_{xy}$, indicating a FQH state at
$\nu=1/4$. The measured value of the plateau, $R_{xy}=3.9h/e^2$, is slightly below the expected
value of $4h/e^{2}$. This is readily apparent in Fig.~\ref{fig:Fig2}, where we have magnified the
regime surrounding $\nu=1/4$. We believe the reduced value of the plateau is due to the mixture of
$R_{xx}$ into $R_{xy}$. In general, the mixing effect can be eliminated by averaging the data from
both positive and negative values of $B$, but due to the extremely high values of magnetic field
($B\sim45$ T) this was not possible. However as further evidence to support our claim of mixing, we
point out that the a reduced Hall resistance is also observed at other well developed FQH states
approaching $\nu=1/4$. At $\nu=2/7$ the corresponding plateau in $R_{xy}$ barely achieves the
expected value of $(7/2)h/e^2$ and then dips below this value. At $\nu=3/11$ the value of the
plateau is $R_{xy}=3.64h/e^2$, slightly below the expected value of $(11/3)h/e^2$. Moving to
$\nu=4/15$ and beyond, the discrepancy between the measured and expected values of $R_{xy}$
continues to increase. It is also possible the observed discrepancy in $R_{xy}$ is caused by a
reduction in the current through the sample due to increased contact resistances. The drive current
was controlled by a 100 M$\Omega$ resistor placed in series with the sample. Contact resistances of
$\sim1$ M$\Omega$ would be necessary to create the observed reductions in $R_{xy}$, which, in our
experience with high mobility samples at large magnetic fields, we find unlikely.

\begin{figure}[b]
\resizebox{3.2 in.}{!}{\includegraphics{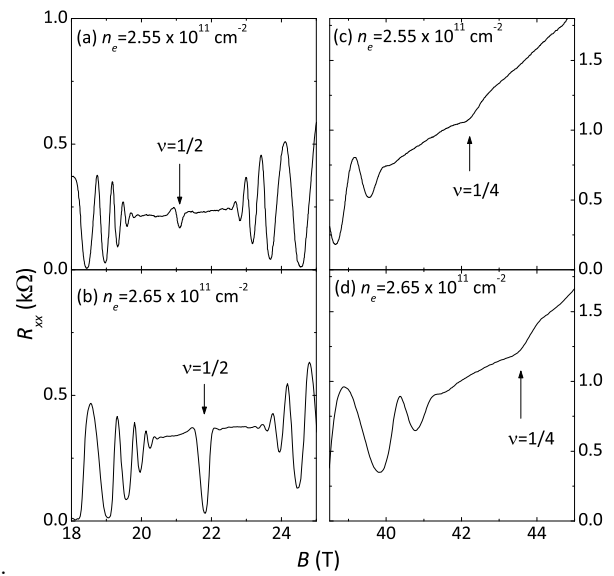}}

\caption{$R_{xx}$ near $\nu=1/2$ and $\nu=1/4$ with $\theta=0^o$ for (a), (c)
$n_e=2.55\times10^{11}$ cm$^{-2}$ and (b), (d) $n_e=2.65\times10^{11}$ cm$^{-2}$. In the case of
$\nu=1/2$ the strength of the FQH state is increased, while for $\nu=1/4$ no FQH state is observed
in either case.}\label{fig:Fig3}
\end{figure}

The data presented here provide strong evidence for the existence of a FQH state at $\nu=1/4$. In
the following, we discuss possible origins of this state. Previous observations of the $\nu=1/2$
state in wide GaAs quantum wells have been discussed\cite{Suen94} within the context of the
quantity $\Delta_{SAS}$, which is the energy difference between the two lowest occupied subbands in
the absence of magnetic field. For a given quantum well, $\Delta_{SAS}$ decreases with increased
density and is thought to decrease with the introduction of parallel magnetic field,
$B_{\parallel}$\cite{Hu,TSLay}. If we assume that $\Delta_{SAS}$ (estimated by calculation as
$\Delta_{SAS}\approx30$ K in our sample at $B=0$) decreases with increased $B_{\parallel}$ then the
increase in the strength of the $\nu=1/2$ state (Fig.~\ref{fig:Fig1}) is qualitatively consistent
with previous observations\cite{Suen94}. In Fig.~\ref{fig:Fig3}(a-b) we show a similar increase in
strength of the $\nu=1/2$ state by increasing the density, again consistent with the decrease in
$\Delta_{SAS}$ and previous measurements. Although not definitive, this provides some evidence that
our $\nu=1/2$ state is similar in nature to previous observations\cite{Suen94} which were found to
have significant overlap with the $\{331\}$ state\cite{SongHe}. However, we also note that in
measurements with a similar density QW with a width of 58 nm and a smaller value of
$\Delta_{SAS}\approx17$ K, no features indicating a FQH state were observed at $\nu=1/2$.

The concomitant emergence and strengthening of the $\nu=1/2$ and $\nu=1/4$ states with $\theta$
shown in Fig.~\ref{fig:Fig1} make it tempting to postulate that the origin of the $\nu=1/4$ state
is also a two-component Halperin state such as $\{553\}$ or $\{771\}$. However, \emph{unlike} at
$\nu=1/2$, the $\nu=1/4$ state does not emerge with the increase in density. As shown in
Fig.~\ref{fig:Fig3}, an increase in density from $n_e=2.55\times10^{11}$ cm$^{-2}$ to
$2.65\times10^{11}$ cm$^{-2}$ causes the strength of the $\nu=1/2$ state to increase significantly,
similar to the behavior shown in Fig.~\ref{fig:Fig1} with increasing $\theta$. In contrast the data
surrounding $\nu=1/4$ show very little change with the increase in density, which highlights a
difference between the two states. In addition, the possibility that $\nu=1/4$ state is described
by $\{771\}$ is unlikely considering the bilayer interpretation of the $\{nnm\}$ wavefunctions. For
the $\{771\}$ wavefunction the filling factor for the electrons in each layer would be $\nu^*=1/7$
and typical single-layer 2DESs enter into an insulating phase beyond $\nu=1/5$ at low
temperatures\cite{PanIP}. A $\{771\}$ would imply a 1/7 FQH state in each layer which seems
unlikely.

Further, for the tilted case, the magnetic length due to the parallel component of magnetic field
at $\nu=1/4$ and $\theta=20.3^o$, given by $\ell_{\parallel}=(\hbar/eB_{\parallel})^{1/2}=6.5$ nm,
is much smaller than the quantum well width $L=50$ nm. Considering this, it is difficult to imagine
that the wave function is not significantly altered by $B_{\parallel}$.

An alternate possible description for the observed $\nu=1/4$ state may be the pairing of composite
fermions, similar to the proposal for a $\nu=1/2$ FQH state in a thick 2DES\cite{park}. The
composite fermions in the $\nu=1/4$ case would consist of one electron with four flux quanta
($^4$CFs). The very large value of magnetic field at the observed $\nu=1/4$ state results in a
relatively decreased value of the magnetic length $\ell_{B}=(\hbar/eB)^{1/2}=3.8$ nm, which
effectively increases the 2DES thickness and thus reduces the short-ranged interaction\cite{park}
and may facilitate the pairing of $^4$CFs.

We also remark on the similarities in the sequences of FQH states between the traces in
Fig.~\ref{fig:Fig1} and that from another high quality GaAs quantum well of the same
width\cite{PanIP}. In Ref.~\cite{PanIP}, $R_{xx}$ is nearly flat around $\nu=1/2$ and $\nu=1/4$
with no observed anomalies when tilted, in contrast to the data presented here. The density of the
sample in Ref.~\cite{PanIP} was $n_e=1\times10^{11}$ cm$^{-2}$, compared to
$n_e\approx2.55\times10^{11}$ cm$^{-2}$ for the data described here. Consequently, $\nu=1/2$ and
$\nu=1/4$ occur at larger values of $B$ in our present sample and therefore the value of $\ell_{B}$
is smaller, which in turn may aid the the formation of a FQH state at $\nu=1/2$ and $\nu=1/4$ as
discussed above.

Finally we make an important point when considering WSQWs. For a given density, it is only in the
limit of very wide wells where electronic behavior is purely bilayer. As the width of the well
decreases the overall behavior is neither entirely consistent with bilayer nor single layer
properties. In the inset of Fig.~\ref{fig:Fig1} we have shown the self-consistently calculated
(excluding the exchange term) electronic band edge and electronic distribution function, $\rho$,
for our sample in the absence of magnetic field. While $\rho$ does have two clear peaks, it is
difficult to characterize it as purely a bilayer distribution.

In conclusion, we report the observation of an even-denominator fractional quantum Hall state at
$\nu=1/4$ in a high quality GaAs quantum well of width $L=50$ nm. The strength of the state
increases with increased tilt angle. Several options relating to the origin of the state are
discussed.

\begin{acknowledgments}
We would like to thank G. Jones, E. Palm, T. Murphy, D. Freeman, and J. Pucci for experimental
assistance. A portion of this work was performed at the National High Magnetic Field Laboratory,
which is supported by NSF Cooperative Agreement No. DMR-0084173, by the State of Florida, and by
the DOE. WP was supported by the DOE/BES at Sandia, a multiprogram laboratory operated by Sandia
Corporation, a Lockheed Martin company, for the United States Department of Energy's National
Nuclear Security Administration under contract DE-AC04-94AL85000. The work at Princeton University
was funded by DOE grant No. DE-FG-02-98ER45683.
\end{acknowledgments}

\end{document}